\documentclass{article}
\usepackage{amsmath,graphics,epsfig,amssymb,eucal}

\newtheorem{theorem}{Theorem}[section]
\newtheorem{lemma}[theorem]{Lemma}
\newtheorem{cor}[theorem]{Corollary}

\newtheorem{definition}{Definition}[section]

\numberwithin{equation}{section}

\newcommand{\weglassen}[1]{}

\renewcommand{\imath}{\mathrm{i}}

\begin{document}
\title{Wannier functions for quasi-periodic finite-gap
potentials}

\author{E.D. Belokolos\thanks{Institute of Magnetism,
Vernadski str. 36, Kiev-142, Ukraine.
}, V.Z. Enolskii\thanks{Department of Mathematics and Statistics,
Concordia University, 7141 Sherbrooke West, Montreal H4B 1R6,
Quebec, Canada.
}, M. Salerno \thanks{Dipartimento di Fisica e Istituto Nazionale di
Fisica della Materia (INFM), via S.Allende, I84081 Baronissi (SA),
Italy.
}}

\date{\empty}

\maketitle

\begin{abstract}
In this paper we consider  Wannier functions of quasi-periodic g-gap
($g\geq 1$) potentials and investigate their main properties. In
particular, we discuss the problem of averaging underlying the
definition of Wannier functions for both periodic and quasi-periodic
potentials  and express Bloch functions and quasi-momenta in terms
of hyperelliptic $\sigma$ functions. Using this approach we derive a
power series expansion of the Wannier function for quasi-periodic
potentials valid at  $|x|\simeq 0$ and an asymptotic expansion valid
at large distance. These functions are important for a number of
applied problems.
\end{abstract}

\section{Introduction}
The Schr\"odinger operator with periodic potentials is
characterized by two distinguished complete sets of orthogonal
functions: the Floquet-Bloch functions, introduced by Floquet
(1882) and F.Bloch (1928), and Wannier functions, introduced by
G.Wannier (1937). The Schr\"odinger operator with algebraically
integrable potentials allows one to construct Bloch (
Floquet-Bloch in one dimension) and Wannier functions and describe
their properties with unprecedented completeness. While Bloch
functions were studied for a long time and there are many reviews
and monographs on the topic, the study of the Wannier functions
corresponding to finite-gap potentials has begun only recently
\cite{bes04}.

The aim of the paper is to further expand  these studies by
presenting a definition and a complete description of the properties
of Wannier functions for $g$-gap potentials with $g\geq 1$. To this
regard, we remark that the usual Wannier functions considered in
Condensed Matter Physics are defined for {\it periodic} potentials.
By considering Wannier functions for $g$-gap potentials, which are
in general {\it quasi-periodic} functions, we are {\it generalizing}
the usual Wannier functions to quasi crystals i.e. to a quasi
periodic lattices,  this being a first main result of the paper.
This circumstance leads to another interesting problem. As well
known, in order to define Wannier functions one must use an
averaging procedure which differs essentially for the periodic and
quasi-periodic cases. Although the averaging problem has been
considered for some algebraically integrable system (see e.g.
\cite{ffm80}, \cite{efmm87}, \cite{kr88}), to the best of our
knowledge, it has not been discussed with sufficient completeness
and not linked to Wannier functions, this being indeed a second main
result of the paper. The study of the averaging problem in
connection with integrable systems, however,  goes beyond the frames
of this paper. Here we shall only remind that averaging procedures
also arise in Krylov-Bogoliubov-Mitropolsky perturbation theory of
integrable systems, in the study of Seiberg-Witten theory, in
multi-matrix models etc. We hope to expand this direction of our
study in the future.

The theory of hyperelliptic curves of different genera leads to a
rich structure of objects which inherits the rich structure of the
moduli spaces of the corresponding hyperelliptic curves. For
example, in the case of elliptic curves $g=1$, periodic and ergodic
cases coincide because there exists only one frequency. In the case
of the two-gap potential the spectral variety is given by the
ultra-elliptic $g=2$ curve and periodic and ergodic cases are rather
different. Moreover, periodic case admits further specialization to
the elliptic periodicity. In the cases $g\geq 3$ a new phenomenon
appears: the curve admits so-called singular half-periods which
leads to certain complications of the theory.

Technically speaking, our development is based on the realization of
hyperelliptic functions in terms of multi-variable
$\sigma$-function. This realization represents a natural
generalization of the Weierstrass theory of elliptic functions to
hyperelliptic functions of higher genera. Higher genera theory was
developed by K.Weierstrass and F.Klein and its exposition is fixed
in classical monograph of H.Baker \cite{ba97}; recent results in the
area are given in \cite{bel97a,bel97b,bel99,bl02} whilst various
applications see e.g. in \cite{eg-rm00}, \cite{on02},\cite{ma02},
\cite{maton03}, \cite{eep03},\cite{balgib03},\cite{balgib04}.

The restricted length of the paper does not permit us to give proofs
of the theorems for which we plan a separate publication.

The paper is organized as follows. In the Section 2 we describe
quasi-periodic potentials from the pure spectral point of view. We
show that the basic objects like finite-gap potential, Weyl and
Bloch functions, which are defined initially on the complex plane
can be reasonably lift to hyperelliptic Riemann surfaces whose
genus coincides with the number of gaps in the spectrum. In the
Section 3 we collect necessary results from the theory of
hyperelliptic $\sigma$-function such as hyperelliptic $\wp$ and
$\zeta$-functions and differential relations between then, as well
as, addition theorems. We also describe certain sub-varieties of
the $\theta$-divisor. In Section 4 we show how to evaluate Bloch
function and quasi-momentum in terms of multidimensional $\sigma$
functions. In section 5 we discuss periodic and quasi-periodic
potentials and show how to  compute averages of squared Bloch
functions in both cases. In Section 6 we discuss Wannier functions
for periodic and quasi-periodic potentials and derive a power
series expansion of the Wannier function for quasi-periodic
potentials valid at  $|x|\simeq 0$ and an asymptotic expansion
valid  at large distance. Finally in Section 7 the main conclusion
and perspective of future work are presented.

\section{The Schr\"odinger operator with quasi-periodic
and finite-gap potential}
\subsection{Quasi-periodic potential}
The function $u(x)$ from a Banach space $C_b(\mathbb{R})$ of
continuous bounded functions is called almost periodic if the set
$\{ T_x u(\cdot ),\; x\in \mathbb{R}\}$, where $T_x u(\cdot
)=u(\cdot +x)$, is relatively compact in $C_b(\mathbb{R})$. A
closure $\Gamma$ of this set is a compact in metrizable Abelian
group. On the set $\Gamma$ there exists a normalized Haar measure
$\mu$ which turns out to be $T_x$--invariant and ergodic. Thus, each
almost--periodic function generates a probability space $(\Gamma,
\mu, T_x)$. The operation of averaging on this space is given by
\begin{equation}\langle f(u)\rangle=\lim_{x\to\infty}\frac{1}{x}\int_0^x
f(T_xu)dx=\int_{\Gamma} f(u)\mu (du). \label{8.1.1}
\end{equation}

By means of the differential expression
$$L(u)=-\partial_{xx}+u(x)$$
we define for $u\in\Gamma$ a Schr\"odinger operator in $\mathcal
{L}^2(\mathbb{R} )$, which is essentially self--adjoint. Let
$\lambda\in\mathbb{C}$ and $c(x,\lambda),\; s(x,\lambda)$ denote
solutions of the equation $L\varphi=\lambda\varphi$ with the initial
data $c(0,\lambda)=1,\; c^{\prime}(0,\lambda)=0,\; s(0,\lambda)=0,\;
s^{\prime}(0,\lambda)=1$. The functions $c(x,\lambda), \;
s(x,\lambda)$ are integral functions of order $1/2$ with respect to
$\lambda$ and continuous with respect to all variables
$(x,\lambda)$. The limits
\begin{equation}w_{\pm}(\lambda)\equiv w_{\pm}(\lambda,
u(\cdot))=\mp\lim_{x\to\infty}
\frac{c(x,\lambda)}{s(x,\lambda)} \label{weilfunctions}
\end{equation}
exist and are called the Weyl functions. Besides these, we shall
also employ the Weyl functions $w_{\pm}(x,\lambda)\equiv
w_{\pm}(\lambda, T_xu(\cdot))$. It is well known that all Weyl
functions are holomorphic in
$\mathbb{C}_+=\{\lambda\in\mathbb{C}\;\vert\;{\rm Im}\lambda>0\}$,
map $\mathbb{C}_+\rightarrow\mathbb{C}_+$, and if they have zeros
or poles on the real axis, these could be simple only \cite{cl55}.
By means of the Weyl functions we can define the functions
\begin{equation}\psi(x,\lambda)=c(x,\lambda)\pm w_{\pm}(x,\lambda)s
(x,\lambda)=\exp\left(\pm\int_0^x w_{\pm}(y,\lambda)dy\right),
\label{8.1.2}\end{equation} which, for each
$\lambda\in\mathbb{C}_+$, belong to $\mathcal{
L}^2(\mathbb{R}_{\pm})$, where $\mathbb{R}_+=[0,\infty)$, and
$\mathbb{R}_-=(-\infty, 0)$, respectively. By definition, the
functions $\psi_{\pm}(x,\lambda)$ satisfy the equation
\begin{equation}-\partial_{xx}\psi_{\pm}(x,\lambda)
+(u(x)-\lambda)\psi_\pm(x,\lambda)=0.\end{equation}
By substituting (\ref{8.1.2}) into this equation, we have the
following equation for the Weyl functions:
\begin{equation}\pm\partial_xw_\pm(x,\lambda)+w_\pm^2(\lambda,
x)+\lambda-u(x)=0. \label{8.1.4}\end{equation}
Following
\cite{jm82}, we define the Floquet function
\begin{equation}f(x,\lambda)=\frac12\langle
w_+(x,\lambda)+w_-(x,\lambda)\rangle.\label{8.1.8}\end{equation} Let
$\lambda=\xi+\imath\eta$. The Floquet function has a finite limit
almost everywhere at $\xi\in\mathbb{R},\;\eta\downarrow 0$
\begin{equation}f(\xi+\imath 0,x)=-l(x,\xi)+\imath\pi n(x,\xi),
\label{8.1.9}\end{equation} where $l(\xi)$ is the Lyapunov
exponent, and $n(\xi)$ is the number of states.

The number of states, $n(x,\xi)$, determines the spectrum
$\Sigma(u)$ of the operator $L(u)$: for a.e. $u\in\Gamma$ the
spectrum is a set of growth points of the number of states
\cite{pastur80},
$$\Sigma(u)={\rm supp}(dn).$$
The Lyapunov exponent $l(x,\xi)$ determines an absolutely
continuous spectrum i.e. a closed set of spectral points with zero
Lyapunov exponent,
$$\Sigma_{a.c.}(u)
=\overline{\{\xi\in\mathbb{R}\;\vert\;l(x,\xi)=0\}},$$ where the
closure is with respect to Lebesgue measure
\cite{pastur80,kotani82}. We now proceed to discuss a special
subset of quasi-periodic potentials, the so called finite--gap
potentials (more details see in
\cite{zmnp80},\cite{bbeim94},\cite{gh03}).

\subsection{The finite-gap potential}
Theory of finite-gap potentials of the Schr\"odinger operator has
long history which goes up to Hermite, Halphene, Darboux. Recent
development was stimulated by the {\it soliton theory} and achieved
in the works made at the middle of seventieth by Novikov,  Dubrovin
and Novikov, Its and Matveev and Krichever (see e.g. \cite{zmnp80}
and references therein).

\begin{definition} The almost--periodic function $u(x)$ is called
a {\it finite--gap potential} if the spectrum of the Schr\"odinger
operator $L(u) =-\partial_{xx}+u(x)$ is the union of a finite set of
segments of  Lebesgue (double absolutely continuous) spectrum
$[E_1,E_2]\cup[E_3,E_4]\cup\cdots \cup[E_{2g_1},\infty]$, where
boundaries of the bands are supposed to be real and ordered as:
$E_1< E_2 \ldots, <E_{2g+1}< +\infty$.
\end{definition}
Starting from this definition, we can derive explicit expressions
and basic properties of finite--gap potentials, Weyl and Bloch
functions. We shall show that the adequate language to describe
these object is the theory of hyperelliptic Riemann surfaces. To
this end we fix the Riemann surface $X$ of genus $g$ of the
algebraic curve given by the equation
\begin{equation}
\mu^2=R_{2g+1}(\lambda), \quad
R_{2g+1}=4\prod_{k=1}^{2g+1}(\lambda-E_k).
\end{equation}
We denote the running coordinate of the curve $X$ as
$P=(\lambda,\mu)$ and coordinate of branch points as $(E_k,0)$,
$k=1,\ldots,2g+1$.
\begin{theorem} The Weyl function of the finite-gap potential $u(x)$ is
defined on the Riemann surface $X$ of the hyperelliptic curve
\begin{equation}
\mu^2=R_{2g+1}(\lambda),\quad  R_{2g+1}(\lambda)=
4\prod_{i=1}^{2g+1} (\lambda-E_i) \label{hypcurve}
\end{equation}
by the formula
\begin{equation}
w(x, P)= \frac{1}{2}\left(\frac{\partial_x
S(x, P)}{ S(x, P)}+\frac{\imath\mu}{
S(x, P)}\right),
\label{weylfunction}
\end{equation}
where
\begin{equation}
S(x,\lambda)=\prod_{k=1}^{g}(\lambda-\lambda_{k}(x))\label{rs}
\end{equation}
and points $(\lambda_k(x),\mu_k(x))$, $k=1,\ldots,g$, are distinct
points of the curve $X$ evaluated accordingly with changing of
$x$.
\end{theorem}
For our  further discussion we need the expression of the number
of states, $n(\xi),$ in terms of the function $w_+(x,\xi),$
$$n(\xi)=\frac{1}{\pi}\left\langle
 w_+(x,\xi)\right\rangle=\frac{1}{\pi}\left\langle \mathrm{Im}\;
w_+(x,\xi)\right\rangle=\frac{1}{2\pi}\int^{\xi}\left\langle\frac{1}{
\mathrm{ Im}\; w_+(x,\xi)}\right\rangle\mathrm{d}\xi.$$ For a
finite--gap potential $u(x)$ we have
$$\mathrm{Im}\;w_+(x,\xi)=\frac{\sqrt{R_{2g+1}(\xi)}}{2S(x,\xi)}.$$
Therefore in this case the number of states is
$$n(\xi)=\frac{1}{\pi}\left\langle
\frac{\sqrt{R_{2g+1}(\xi)}}{2S(x,\xi)}\right\rangle=\frac{1}{\pi}
\int^{\xi}\left\langle\frac{S(x,\xi)}{\sqrt{R_{2g+1}(\xi)}}\right\rangle\mathrm{d}\xi.$$
The number of states and the wave number for all finite-gap
potentials are connected as follows
$$n(\xi)=\frac{1}{\pi}k(\xi).$$ A direct consequence of the above
discussion is the following theorem.
\begin{theorem}
The quasi-momentum of eigenfunction can be defined by the expression
\begin{align}
k(P)=\langle  w(x,P) \rangle = \left\langle \frac{ \mu }{S(x,P)}
\right\rangle =\int\limits\limits_{P_0}^P \frac{\langle S(x,P)
\rangle} {\mu}\mathrm{d}\lambda,\label{quasimomentum}
\end{align}
where $P=(\lambda,\mu)$ and $P_0=(\lambda_0,\mu_0)$ are points of
the hyperelliptic curve $\mu^2=R_{2g+1}(\lambda)$.
\end{theorem}
The last integral in Eq. (\ref{quasimomentum}) is the
Schwarz-Cristoffel integral which maps the l.h.s. half-plane of
complex $\lambda$-plane, to a rectangle in the complex $k$-plane
\cite{marost75,gmn80}. We shall further interpret this integral as
a second kind Abelian integral on $X$ over meromorphic
differential,
\begin{equation}
\mathrm{d}\mathcal{K}=\frac{\langle S(x,P) \rangle}
{\mu}\mathrm{d}\lambda,\label{diffquasimom}
\end{equation}
which we shall call differential of quasi-momentum.
\begin{theorem}
The eigenfunction of the $g-$gap potential, which
is normalized by the condition
\begin{equation}\langle|\psi(x,P)|^2\rangle=1,\label{psinorm}
\end{equation}
is of the form
\begin{equation}
\psi(x,P)=\sqrt{\frac{S(x,P)}{\langle S(x,P)\rangle}
}\exp\left\{\frac{\imath \mu }{2}\int\limits_{x_0}^{x}
\frac{\mathrm{d}y}{ S(y,P)}\mathrm{d}y\right\}.
\label{psiricatti}
\end{equation}
\end{theorem}
We see that in order to define the quasi-momentum and normalize
the eigenfunction we must fulfill the averaging. As a consequence
we should say some words about the averaging for quasi- and
almost-periodic functions since the potentials with finite number
of gaps are quasi-periodic functions and the potentials with
infinite number of gaps are almost-periodic functions.
\subsection{Averaging for the finite-gap potentials:
periodic and ergodic cases} Any almost-periodic function has a
mean value
\begin{equation*}
\langle f\rangle =
\lim_{L\to\infty}\frac{1}{L}\int_{0}^{L}f(x)\mathrm{d}x.
\end{equation*}
This allows for any almost-periodic function to build a Fourier
series
\begin{equation*} f(x) \simeq \sum_{n} A_n
\exp(\imath\lambda_n x),\quad A_n = \langle
f(x)\exp(\imath\lambda_n x)\rangle.
\end{equation*}
The numbers $\lambda_n$ are designated as the Fourier
frequencies and the numbers $A_n$ as the Fourier coefficients of
the function $f(x)$.

We remind that a countable set of real numbers
$\lambda_1,\ldots,\lambda_n,\ldots$ has a rational basis
$\alpha_1, \alpha_2, \ldots$ if the numbers  $\alpha_1, \alpha_2,
\ldots$ are linearly independent and any number $\lambda_n$ can be
presented as their finite linear combination with rational
coefficients, i.e.
\begin{equation*} \lambda_n =
\sum_{k=1}^{S_n}r_{k}^{n}\alpha_{k},\quad r_{k}^{n} \in
\mathbb{Q}.\end{equation*}
We say that the basis is finite if it
is finite set, we say that the basis is integer if all numbers
$r_{k}^{n}$ are integer numbers.  If the Fourier frequencies of
almost-periodic function have a finite and integer basis we
designate the appropriate almost-periodic function as the
quasi-periodic one. A quasi-periodic function with one period is
pure periodic one.

Function $F(x_1, x_2,\ldots)$ of finite or countable set of
variables, each of which admits all real values, is called
limiting periodic if it is a uniform limit of periodic ones, i.e.
if for any positive real number $\varepsilon$ we can point out
such an integer positive number $n(\varepsilon)$ and such a
periodic function $F_{\varepsilon}(x_1, x_2,\ldots, x_n)$ that
\[\sup_{-\infty< x_k, k=1,2,\ldots < +\infty} |F(x_1, x_2, \ldots) -
F_{\varepsilon}(x_1, x_2, \ldots, x_n)| < \varepsilon .\] It
appears that for any almost-periodic function $f(x)$ there exists
such a limiting periodic function $F(x_1, x_2, \ldots)$ of finite
or countable set of variables such that
\begin{equation*}f(x) = F(x, x, \ldots).\end{equation*}
In other words any almost-periodic function is a diagonal
restriction  of some limiting periodic function. The properties of
the limiting periodic function  $F(x_1, x_2, \ldots)$ depend
essentially on the basis of the Fourier frequencies of the
function $f(x)$. If the basis $\alpha_1, \alpha_2, \ldots$ of the
almost-periodic function $f(x)$ is integer then the limiting
periodic function  $F(x_1, x_2, \ldots)$ is periodic with periods
$2\pi/\alpha_1, 2\pi/\alpha_2, \ldots.$ If the basis $\alpha_1,
\alpha_2, \ldots$ of the almost-periodic function $f(x)$ is finite
then the limiting periodic function $F(x_1, x_2, \ldots)$ depends
on finite set of variables.  If the basis $\alpha_1, \alpha_2,
\ldots$ of the almost-periodic function $f(x)$ is finite and
integer then the limiting periodic function $F(x_1, x_2, \ldots)$
is a periodic function on finite set of variables. Obviously, the
mean value of the function $f(x)$  depends essentially on whether
this function is periodic, quasi-periodic or almost-periodic. If
the function $f(x)$ is periodic then the mean value of the
function is just the average over the period. If the function
$f(x)$ is quasi-periodic or almost-periodic then the mean value of
the function is the average over the phase space, i.e. ergodicity
takes place. This is illustrated with the following classical
statement.
\begin{theorem}[\bf{Kronecker--Weyl}] Let $\lambda_k, k=1,\ldots,n$ be
real linearly independent numbers, $\theta_k, k=1,\ldots,n$ be
arbitrary real numbers, $\delta_k, k=1,\ldots,n$ be arbitrary
positive numbers. Let $\chi(x_1,x_2,\ldots,x_n)$ be a
characteristic function of parallelepiped in $\mathbb{R}^n$
defined by inequalities
\[\theta_k-\delta_k<x_k<\theta_k+\delta_k,\quad k=1,\ldots,n.\]
Then, uniformly in $L$ we have
\begin{equation*}
\lim_{L\to\infty}\int_0^L\chi(\lambda_1x-\theta_1,\ldots,
\lambda_nx-\theta_n)\mathrm{d}x = \pi^{-n} \delta_1\ldots\delta_n,
\end{equation*}
\end{theorem}
where the function $\chi(x_1,x_2,\ldots,x_n)$ is the periodic
continuation to the whole $\mathbb{R}^n$  in all variables $x_k,
k=1,\ldots,n$, with periods $2\pi$.

In the problem under consideration a finite-gap potential appears
as a restriction to a straight line of some meromorphic function
defined on a Jacobian. Of course on this line the potential must
be real and non-singular. Depending on a direction of the line
with respect to periods of the Jacobian this finite-gap potential
can be periodic or quasi-periodic and this circumstance influences
the averaging essentially. The averaging of quasi-periodic
functions which are restrictions of the $\theta-$functions was
previously described in the literature (see e.g. \cite{weyl16}),
and the averaging of solutions of integrable evolution equations
(see e.g.\cite{ffm80},\cite{efmm87}). We shall use these results
when discussing the averaging for the finite-gap potentials. We
first shall formulate our problem in terms of the
$\sigma-$functions.

We remark that it is convenient to use the language of
$\sigma-$function in our foregoing discussions since all
expressions in $\sigma-$functions are form-invariant with respect
to the number of gaps of potential (or the genus of curve) and
this fact is very useful for exposition. Keeping this in mind, we
give in the next section a brief account of necessary facts from
the theory of hyperelliptic $\sigma-$ functions.

\section{Hyperelliptic $\sigma-$functions}
The theory of hyperelliptic $\sigma-$functions represents a
many-dimensional generalization of the theory of elliptic functions
and provide suitable language of our development.
\subsection{Curve and differentials}
\begin{figure}
\begin{center}
\unitlength 0.7mm \linethickness{0.4pt}
\begin{picture}(150.00,80.00)
\put(-11.,33.){\line(1,0){12.}} \put(-11.,33.){\circle*{1}}
\put(1.,33.){\circle*{1}} \put(-10.,29.){\makebox(0,0)[cc]{$E_1$}}
\put(1.,29.){\makebox(0,0)[cc]{$E_2$}}
\put(-5.,33.){\oval(20,30.)}
\put(-12.,17.){\makebox(0,0)[cc]{$\mathfrak{ a}_1$}}
\put(-5.,48.){\vector(1,0){.7}}
\put(12.,33.){\line(1,0){9.}} \put(12.,33.){\circle*{1}}
\put(21.,33.){\circle*{1}} \put(13.,29.){\makebox(0,0)[cc]{$E_3$}}
\put(22.,29.){\makebox(0,0)[cc]{$E_4$}}
\put(17.,33.){\oval(18.,26.)}
\put(10.,19.){\makebox(0,0)[cc]{$\mathfrak{ a}_2$}}
\put(16.,46.){\vector(1,0){.7}}
\put(35.,33.){\circle*{1}} \put(40.,33.){\circle*{1}}
\put(45.,33.){\circle*{1}}
\put(60.,33.){\line(1,0){9.}} \put(60.,33.){\circle*{1}}
\put(69.,33.){\circle*{1}}
\put(60.,29.){\makebox(0,0)[cc]{$E_{2g-1}$}}
\put(72.,29.){\makebox(0,0)[cc]{$E_{2g}$}}
\put(65.,33.){\oval(30,15.0)}
\put(59.,20.){\makebox(0,0)[cc]{$\mathfrak{ a}_g$}}
\put(62.,40.2){\vector(1,0){.7}}
\put(114.,33.00){\line(1,0){33.}} \put(114.,33.){\circle*{1}}
\put(147.,33.){\circle*{1}}
\put(115.,29.){\makebox(0,0)[cc]{$E_{2g+1}$}}
\put(146.,29.){\makebox(0,0)[cc]{$E_{2g+2}=\infty$}}
\put(18.,72.){\makebox(0,0)[cc]{$\mathfrak{ b}_1$}}
\put(25.,78.){\vector(4,1){0.2}}
\bezier{484}(-4.,33.00)(0.,76.)(25.,78.)
\bezier{816}(25.00,78.)(75.00,82.00)(143.00,33.00) 
\put(33.,64.){\makebox(0,0)[cc]{$\mathfrak{ b}_2$}}
\put(38.00,70.){\vector(4,1){0.2}}
\bezier{384}(17.,33.00)(17.,68.)(38.00,70.)
\bezier{516}(38.00,70.)(80.00,74.00)(137.00,33.00) 
\put(72.,48.){\makebox(0,0)[cc]{$\mathfrak{ b}_g$}}
\put(72.,54.){\vector(3,1){0.2}}
\bezier{226}(64.00,33.)(64.00,52.00)(72.,54.)
\bezier{324}(72.,54.)(88.,58.00)(125.00,33.00) 
\end{picture}
\end{center}
\caption{A homology basis on a Riemann surface of the
hyperelliptic curve of genus $g$ with real branching points
$E_1,\ldots,E_{2g+2}=\infty$ (upper sheet).  The cuts are drawn
from $E_{2i-1}$ to $E_{2i}$ for $i=1,\dots,g+1$.  The $\mathfrak
b$-cycles are completed on the lower sheet (the picture on lower
sheet is just flipped horizontally).} \label{figure-1}\end{figure}
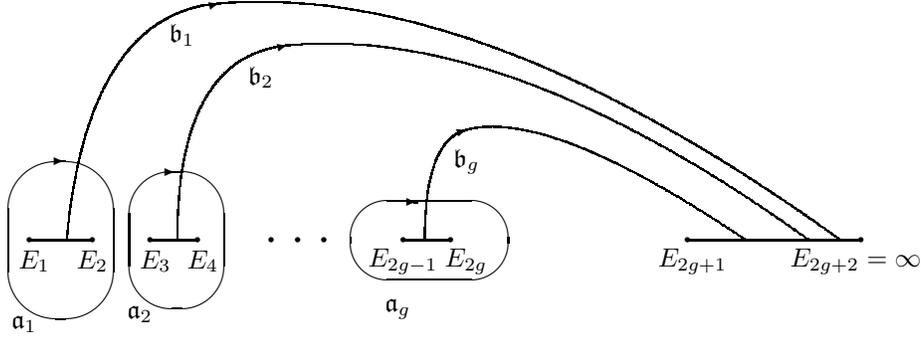

Let $X $  be the hyperelliptic
curve given by the equation
\begin{equation}
 \mu^2= 4\lambda^{2g+1}+\sum_{i=0}^{2g} \alpha_{i}\lambda^{i}
= 4\prod_{k=1}^{2g+1} (\lambda-E_{k}) =R (\lambda),
\label{xcurve}
\end{equation}
realized as a two sheeted covering over the Riemann sphere
branched in the real points $(E_k,0)$, $k \in {\mathcal G} =
\{1,\ldots,2g+1\}$, with $E_j\neq E_k$ for $j\neq k$, and at
infinity, $E_{2g+2}=\infty$. The order of branch points is
according to $E_1 < E_2 < \ldots < E_{2g+1}$
(see~Fig.~\ref{figure-1}).

A set of $2g$ canonic holomorphic differentials
$\mathrm{d}\boldsymbol{h}=(\mathrm{d}h_1,\ldots,\mathrm{d}h_g )^T$ and
meromorphic differentials
$\mathrm{d}\boldsymbol{r}=(\mathrm{d}r_1,\ldots,\mathrm{d}r_g )^T$ is
defined by the following expressions,
\begin{align}
\mathrm{d}h_j&=\frac{\lambda^{j-1}}{\mu}\mathrm{d}\lambda,\quad
j=1,\ldots,g,
\label{holomorphic}\\
\mathrm{d}r_j&=\sum_{k=j}^{2g+1-j}(k+1-j)\alpha_{k+1+j}
\frac{\lambda^k}{4\mu} \mathrm{d}\lambda,\quad j=1,\ldots,g.
\label{meromorphic}
\end{align}
We denote periods
\begin{align}\begin{split}
2\omega&= \left( \oint_{\mathfrak{a}_k}
\mathrm{d}h_i\right)_{i,k=1,\ldots,g},\quad 2\omega'=
\left( \oint_{\mathfrak{b}_k} \mathrm{d}h_i\right)_{i,k=1,\ldots,g},\\
2\eta&= \left( -\oint_{\mathfrak{a}_k}
\mathrm{d}r_i\right)_{i,k=1,\ldots,g},\quad 2\eta'= \left(
-\oint_{\mathfrak{b}_k}
\mathrm{d}r_i\right)_{i,k=1,\ldots,g}.\end{split} \label{periods}
\end{align}

The half-periods  $\omega,\omega',\eta,\eta'$ satisfy the
generalized Legendre relation
\begin{equation}
{\mathfrak M}\left(\begin{array}{cc}0&-1_g\\
1_g&0\end{array}\right)
{\mathfrak M}^T=
\frac{\imath\pi}{2}\left(\begin{array}{cc}0&-1_g\\
1_g&0\end{array}\right),\label{legendre}
\end{equation}
where $1_g$ is the unit $g\times g$ matrix and $\mathfrak M$ is
the $2g\times 2g$-matrix $$\mathfrak M=\left(
\begin{array}{cc}\omega&\omega'\\\eta&\eta'\end{array}
\right).$$ Denote $\tau={\omega}^{-1}\omega'$-matrix belonging to
the Siegel upper half-space $\mathcal{S}_g=\{\tau| \tau^T=\tau, \;
\mathrm{Im}\; \tau >0  \}$ and necessarily symmetric matrix
$\varkappa= \eta(2\omega)^{-1}$.  Jacobi variety of the curve is
defines as the factor over the normalized period lattice,
$\mathrm{Jac}(X)=\mathbb{C}/ 1_g\oplus\tau $. Abel map
$\boldsymbol{\mathfrak{A}}:\{ P\in X\} \rightarrow \mathrm{Jac}(X) $
maps the point $P$ of the curve to Jacobian by the rule
$$\boldsymbol{\mathfrak{A}}(P)=\int_{P_0}^{P}\mathrm{d}\boldsymbol{v},  $$
where $P_0$ is the base point of the Abel map. It maps a positive
divisor $D$ of degree $n$,   $D=P_1+\ldots+P_n$ as follows
\[
\boldsymbol{\mathfrak{A}}(D)=
\boldsymbol{\mathfrak{A}}(P_1)+\ldots
+\boldsymbol{\mathfrak{A}}(P_n).
\]
In what follows we deal with non-special divisors (we remind that a
special divisor $D$ is a divisor for which it exists a meromorphic
function on $X$ with poles at most in $D$). The base point of the
Abel map will be taken as $P_0=(\infty,\infty)$ unless a contrary is
stated.

\subsection{$\theta$-functions and $\sigma$-functions}
Introduce canonic $\theta$-function,
$\theta(\boldsymbol{z};\tau),$ by the formula
\begin{equation}
\theta(\boldsymbol{z};\tau)=\sum_{\boldsymbol{m}\in\mathbb{Z}^g}
\mathrm{exp}\left\{\imath\pi \boldsymbol{m}^T\tau\boldsymbol{m}
+2\imath\pi \boldsymbol{z}^T\boldsymbol{n} \right\}.\label{theta}
\end{equation}
Fundamental Riemann vanishing theorem says that if for a vector
$\boldsymbol{z}\in\mathrm{Jac}(X)$ one has that
$\theta(\boldsymbol{z};\tau)\not\equiv0$, then there exists a
non-special divisor $D$ of degree $g$ and a vector
$\boldsymbol{K}_{\infty}$ (vector of Riemann constant with the base
point $P_0=(\infty,\infty)$) such that
\[ \boldsymbol{z}=\boldsymbol{\mathfrak{A}}(D)-\boldsymbol{K}_{\infty}.\]
Moreover Riemann $\theta$-function
\[ \theta(P)=\theta(\boldsymbol{\mathfrak{A}}(P)
-\boldsymbol{\mathfrak{A}}(D)+\boldsymbol{K}_{\infty};\tau  )\]
vanishes precisely in $g$ points of the divisor $D$. Let
$(\lambda_1,\mu_1),\ldots, (\lambda_g,\mu_g)$ be a non-special
divisor. Denote
\begin{equation}
\boldsymbol{h}=\sum_{k=1}^g\int_{\infty}^{\lambda_k} \mathrm{d} u_k
-\boldsymbol{K}_{\infty}.
\end{equation}
Then the $\sigma$-function is defined by the formula
\begin{align}
\sigma(\boldsymbol{h})&=\sqrt{\frac{\pi^g}{\det
(2\omega)}}\sqrt[4]{\prod_{1\leq i < j \leq 2g+1} (E_i-E_j)}
\notag\\
&\times\theta((2\omega)^{-1}\boldsymbol{h}|\tau
)\mathrm{exp}(\boldsymbol{h}^T\eta(2\omega)^{-1}\boldsymbol{h};\tau
).\label{kleinsigma}
\end{align}
The function (\ref{kleinsigma}) represents a $g-$dimensional
generalization of the elliptic $\sigma$-function. It is invariant
with respect to the action of certain subgroups of the symplectic
group. It also has the following transformation property with
respect to shift of the argument by a period,
\begin{equation}
\sigma(\boldsymbol{h}+2\omega\boldsymbol{ n}+ 2\omega'\boldsymbol{m} )
=\sigma(\boldsymbol{h}) \mathrm{exp}
\left\{ (2\eta\boldsymbol{n}+2\eta'\boldsymbol{m})^T(\boldsymbol{h}
+\omega\boldsymbol{n}+\omega'\boldsymbol{m}  )  \right\},
\label{transfsigma}
\end{equation}
where $\boldsymbol{n}, \boldsymbol{m}$ are integer $g$-vectors.

Variety of zeros of the $\sigma$-function called  $\sigma$-divisor
$\Theta$, coincides with the $\theta$-divisor $(\theta)$. The
$\sigma$-divisor in rational limit is given as zeros of a {\it
Schur-Weierstrass polynomial} analyzed in \cite{bel99}.

Denote Abelian image $\Theta_1$ of the curve $X$,
\[ \Theta_1=
\left\{\boldsymbol{v}\in \mathrm{Jac}(X) |\quad
\boldsymbol{v}=\int_{\infty}^P\mathrm{d}\boldsymbol{h},\quad P\in X
\right\} .\] At genera $g>1$  the variety  $\Theta_1$ belongs to the
$\sigma$-divisor, $\Theta_1\subset\Theta$. We shall need the
following result from \cite{on04} describing vanishing of certain
partial derivatives of the $\sigma$-function.

\begin{lemma} [{\bf {\^O}nishi, 2004}]\label{onishi}
Let $\sigma_{\mathcal{I}}$ be partial derivative
\[ \sigma_{\mathcal{I}}(\boldsymbol{h})
=\prod_{j\in \mathcal{I} } \frac{\partial}{\partial h_j}  \sigma
(\boldsymbol{h}),   \] where $\mathcal{I}$ be multi-index given for
different genera $g$ of hyperelliptic curve $X$ in the Table 1
\begin{center}
\begin{tabular}{|c|c|c|c|c|c|c|c|c|c|c|}
\hline
$g$&1&2&3&4&5&6&7&8&9&\ldots\\
\hline
$\mathcal{I}$&&2&2&24&24&246&246&2468&2468&\ldots\\
\hline
\end{tabular}
\end{center}
\begin{center} Table 1.  Multi-indices corresponding to different genera $g$
\end{center}

Then in the vicinity of the point
$\boldsymbol{h}=(0,\ldots,0)^T\in\Theta_1$ the following expansions
are valid for the $\lambda$-coordinate of the curve $X$ and partial
derivative $\sigma_{\mathcal{J}}$
\begin{align}
\lambda
=\frac{1}{\xi^2}+O(1),\\
\sigma_{\mathcal{I}}(\boldsymbol{h})^2=\xi^{2g}+ O(\xi^{2g+2}).
\end{align}
Moreover all derivatives $ \sigma_{\mathcal{J}}(\boldsymbol{h})$
vanish on $\boldsymbol{h}\in\Theta_1$ for all multi-indices
$\mathcal{J}\subset\mathcal{I}$, $|\mathcal{J}| < |\mathcal{I}|.$
\end{lemma}
The proof of the lemma is based on the detailed analysis of the
Schur polynomials associated with hyperelliptic curves and certain
results of \cite{bel99}.

\subsection{Kleinian $\wp$ and $\zeta$-functions}
Introduce $g-$dimensional $\zeta$ and $\wp$-functions as logarithmic
derivatives of the $\sigma$-function as follows
\begin{align*}
\zeta_i(\boldsymbol{h})&= \frac{\partial}{\partial u_i }\,
\mathrm{log}\,\sigma(\boldsymbol{h}),\quad i=1\ldots, g,\\
 \wp_{i,j}(\boldsymbol{h})&=-\frac{\partial^2}{\partial u_i\partial u_j
}\, \mathrm{log}\, \sigma(\boldsymbol{h}),\quad  i,j=1,\ldots, g,\\
\wp_{i,j,k}(\boldsymbol{h})&=-\frac{\partial^3}{\partial u_i\partial
u_j \partial u_k}\, \mathrm{log}\, \sigma(\boldsymbol{h}), \quad
i,j,k=1,\ldots, g, \quad \text{etc.}
\end{align*}
These functions have the following periodicity properties. Let
$\boldsymbol{n}, \boldsymbol{m}\in\mathbb{Z}^g$ two arbitrary integer
vectors. Then \begin{align}
&\zeta_{k}(\boldsymbol{u}+2\omega\boldsymbol{n}+2\omega'\boldsymbol{m}  )
=\zeta_{k}(\boldsymbol{u})+2\eta\boldsymbol{n}+2\eta'\boldsymbol{m}, \quad
k=1,\ldots, g,\label{perzeta}\\
&\wp_{\mathcal{I}}(\boldsymbol{u}+2\omega\boldsymbol{n}+2\omega'\boldsymbol{m}
)= \wp_{\mathcal{I}}(\boldsymbol{u}),\label{perwp}
\end{align}
where $\mathcal{I}$ be arbitrary set of two, three, etc., indices.

Consider the principal object of the hyperelliptic theory the {\it
master polynomial}
\[ F(\lambda;\lambda_1,\ldots,\lambda_g)=\prod_{k=1}^{g}(\lambda-\lambda_k), \]
where $\lambda_1,\ldots,\lambda_g$ are first coordinated of
non-special divisor of degree $g$, $P_1=(\lambda_1,\mu_1),\ldots,
P_g=(\lambda_g,\mu_g)$. O.Bolza \cite{bo95}  found expression for
the master polynomial in terms of $\wp$-functions (se also
\cite{ba97},\cite{ba98},\cite{bel97a} for derivation and
applications) in the form
\begin{equation} F(\lambda;\lambda_1,\ldots,\lambda_g)\equiv
\mathcal{P}(\lambda,\boldsymbol{h} ) =\lambda^g-
\wp_{g,g}(\boldsymbol{h})\lambda^{g-1}-\ldots-\wp_{g,1}
(\boldsymbol{h}).\label{JIP1}
\end{equation}
In what follows we shall refer to this parametrization of the master
polynomial as to {\it Bolza polynomial}.

Zeros of  Bolza polynomial $\mathcal{P}(\lambda,\boldsymbol{h} )$
are known to solve the Jacobi inversion problem,
\begin{align}\begin{split}
&\lambda_1+\ldots+\lambda_g=\wp_{g,g}(\boldsymbol{h}),\\
&\qquad\qquad\vdots\\
&\lambda_1\cdots\lambda_g=(-1)^{g+1}\wp_{1,g}(\boldsymbol{h}),\end{split}\label{jip}
\end{align}
whilst the $\mu$-coordinates of the divisor
$P_1=(\mu_1,\lambda_1),\ldots ,  P_g=(\mu_g,\lambda_g)$ are given by
\begin{equation}
\mu_k=-\left.\frac{\partial}{\partial u_g}
\mathcal{P}(\lambda,\boldsymbol{h})\right|_{\lambda=\lambda_k},
\quad k=1,\ldots,g. \label{JIP2}
\end{equation}

Among various relations of the theory we shall use, for our
derivation the following formula for $\zeta$-function,
\begin{equation}
-\int\limits_{E_{2g+1}}^P\frac{\lambda^g}{\mu}\mathrm{d}
\lambda
+\frac12\frac{\mu}{\mathcal{P}(\lambda,\boldsymbol{h})}
+\frac12\frac{\partial_g\mathcal{P}(\lambda,\boldsymbol{h})}
{\mathcal{P}(\lambda,\boldsymbol{h})}
=\zeta_g\left(\boldsymbol{h}+\int\limits_{\infty}^P
\mathrm{d}\boldsymbol{h}\right)
-\zeta_g(\boldsymbol{h}). \label{zetaformula}
\end{equation}
The derivation of this formula is documented in \cite{ba97,ba98} and
\cite{bel97a,bel97b} and based on the finding of explicit expression
for the fundamental bi-differential - so called {\it Bergman kernel}
in modern terminology.

Another set of formulae describe Jacobi variety $\mathrm{Jac}(X)$ as
algebraic variety in $\mathbb{C}^{g+\frac{g(g+1)}{2}}$. The
functions $\wp_{ggi}$ and $\wp_{ik}$ are related by
\begin{align}\begin{split}
\lefteqn{  \wp_{ggi}  \wp_{ggk}=4  \wp_{gg}  \wp_{gi}  \wp_{gk}-
2 (  \wp_{gi}  \wp_{g-1, k}+  \wp_{g, k}  \wp_{g-1, i}) }   \\
&+ 4 (  \wp_{gk}  \wp_{g, i-1}+  \wp_{gi}  \wp_{g, k-1}) +
4  \wp_{k-1, i-1}-2 (  \wp_{k, i-2}+  \wp_{i, k-2})    \\
&+ \alpha_{2g}  \wp_{gk}  \wp_{gi}+  \frac{  \alpha_{2g-1}}2 (
\delta_{ig}  \wp_{kg}+  \delta_{kg}  \wp_{ig}) +c_{i, k
},\end{split} \label{product3}
\end{align}
where
\begin{equation}c_{ i, k }=  \alpha_{2i-2}  \delta_{ik}
+  \frac12 (  \alpha_{2i-1}  \delta_{k, i+1} +  \alpha_{2k-1}
\delta_{i, k+1}). \label{cij}
\end{equation}
We shall also use derivative of these formulae,
\begin{equation}
\wp_{gggi}=(6\wp_{gg}+\alpha_{2g})\wp_{gi}+6\wp_{g,i-1}-2\wp_{g-1,i}
+\frac12\delta_{g,i}\alpha_{2g-1}, \label{kdv}
\end{equation}
which describe according to \cite{bel97a} the KdV hierarchy. We
remark that the above formulae were derived for the case
ultraelliptic curves $g=2$ in \cite{ba07} where relation to
integrable equations like KdV or sine-Gordon had not been noticed.
The comprehensive generalization to hyperelliptic curves of higher
genera was obtained to the first time in \cite{bel97a,bel97b}.

We complete our brief description of $\sigma$-functions by the
addition theorem {\it point+divisor} for hyperelliptic
$\sigma$-functions.
\begin{theorem} \label{addlemma} Let $$\boldsymbol{v}=\int_{(\infty,\infty)}^{(\lambda,\mu)}
\mathrm{d}\boldsymbol{h}\in
\Theta_1$$ be Abelian image of the curve $X$ and let $
\boldsymbol{h}$ be a vector in general position in
$\mathrm{Jac}(X)$. Then the following formula is valid
\begin{equation}
\frac{\sigma(\boldsymbol{h}-\boldsymbol{v})
\sigma(\boldsymbol{h}+\boldsymbol{v})   }
{\sigma_{\mathcal{I}}(\boldsymbol{v})^2\sigma(\boldsymbol{h})^2}
=\mathcal{P}(\lambda,\boldsymbol{h}), \quad \label{divadd}
\end{equation}
where $\mathcal{P}(\lambda,\boldsymbol{h})$ is the Bolza polynomial
(\ref{JIP1}) and the multi-index $\mathcal{I}$ is given in the Table
of Lemma \ref{onishi}.
\end{theorem}

To the best knowledge of the authors the formula (\ref{divadd}) as
well the foregoing relation (\ref{rel22}) are new. We shall publish
detailed proof elsewhere and give here only its short version. This
proof is based on the Riemann vanishing theorem and solution of the
Jacobi inversion problem in the form (\ref{jip}) what leads to the
Bolza polynomial in the right hand side. Lemma \ref{onishi} permits
to check the leading terms of the expansions in $\boldsymbol{v}$
near the point $\boldsymbol{v}=(0,\ldots,0)^T\in\Theta_1$ in both
sides of the equality.

\begin{cor}\label{quotlemma}
Let $\boldsymbol{\Omega}=\sum_{k\in\mathbb{I}}
\boldsymbol{\mathfrak{A}}_k$, where $|\mathbb{I}|=g$ be non-singular
even half-period represented in the form
$\boldsymbol{\Omega}=2\omega \boldsymbol{n}+2\omega'\boldsymbol{n}'$
with $\boldsymbol{n},\boldsymbol{n}'\in\mathbb{Z}^g$.   Then the
following formula is valid
\begin{equation}
\frac{\sigma(\boldsymbol{v}-\boldsymbol{\Omega})^2 }
{\sigma_{\mathcal{I}}(\boldsymbol{v})^2}
=\mathcal{Q}_0(\lambda)\prod_{i\in\mathbb{I}} (\lambda-E_i), \quad
\boldsymbol{v}\in \Theta_1,\quad
\boldsymbol{\Omega}\in\mathrm{Jac}(X),\label{rel22}
\end{equation}
where $\mathcal{Q}_0(\lambda)=\sigma(\boldsymbol{\Omega})^2
\mathrm{exp} \{\boldsymbol{\Delta}^T
\int_{\infty}^P\mathrm{d}\boldsymbol{h} \}$ with
$\boldsymbol{\Delta}=2\eta\boldsymbol{n}+2\eta'\boldsymbol{n}'$,
$E_i$ are branch points of the curve $X$ and $\mathcal{I}$ is the
set of multi-indices given in the Lemma \ref{onishi}
\end{cor}

\section{$\sigma$-functional realization of the basic functions}
Consider $\boldsymbol{h}$-vectors of the form
\begin{equation}
\boldsymbol{h}=\imath x \boldsymbol{e}_g-\boldsymbol{\Omega}, \quad
\boldsymbol{e}_g=(0,\ldots,0,1)^T,
\end{equation}
where $\boldsymbol{\Omega}=(\Omega_1,\ldots,\Omega_g)^T$ be {\it non
pure imaginary even non-singular half-period} supported by $g$
branch points
\begin{equation}\boldsymbol{\Omega}
= \sum_{i\in\mathbb{I}}\boldsymbol{\mathfrak{A}}_i,
\quad |\mathbb{I}|=g.\label{omega}
\end{equation}
Then the Bolza polynomial is identified as the $S(x,\lambda)$
function introduced in Section 2,
\begin{equation}
\mathcal{P}(\lambda;\boldsymbol{h}) =S(x,\lambda). \label{bolzas}
\end{equation}
\subsection{Its-Matveev theorem}
The fol\-lowing theorem describes al\-geb\-ro\--geo\-metric solutions of
the Schr\"o\-dinger equation with finite-gap potentials and was proved by
Its and Matveev in 1975 \cite{im75}. Different variants of proof the
Its-Matveev theorem are known see e.g.  \cite{bbeim94}, \cite{du81},
\cite{gh03}.
\begin{theorem} [{\bf Its-Matveev, 1975}] \label{its-matveev}
Let $X$ be non-degenerate hyperelliptic curve of genus $g$. Let
$\boldsymbol{\Omega}=(\Omega_1,\ldots,\Omega_g)^T$ be non pure
imaginary even non-singular half-period supported by $g$ branch
points. Then the smooth and real potential $u(x)$ and  one-valued on
the curve $X$ Bloch function $\psi(x,P)$, are given by the formulae
\begin{equation}
u(x)=\wp_{gg}(\imath
x\boldsymbol{e}_g-\boldsymbol{\Omega}),\label{usigma}
\end{equation}
\begin{align}\begin{split}
\psi(x,P)&=C(P)
\frac{\sigma\left(\int_{\infty}^P\mathrm{d}\boldsymbol{h} -\imath
x\boldsymbol{e}_g+\boldsymbol{\Omega}\right)}
{\sigma_{\mathcal{I}}
\left(\int_{\infty}^P
\mathrm{d}\boldsymbol{h}\right)
\sigma(\imath x\boldsymbol{e}_g -\boldsymbol{\Omega})}\\
\times& \mathrm{exp}\left\{\imath x\int\limits_{(E_{2g+1},0)}^{P}
\mathrm{d}r_g-\boldsymbol{\Omega}^T
\int\limits_{(E_{2g+1},0)}^P\mathrm{d}\boldsymbol{r} \right\},
\end{split}
\label{psisigma}
\end{align}
where $\mathcal{I}$ is the set of indices given in the Table of the
Lemma \ref{onishi},  $C(P)$ is a normalization constant which
depends on the point $P\in X$, $\boldsymbol{e}_g=(0,\ldots,0,1)^T$.
\end{theorem}
\subsection{$\sigma$-functional description of the
 Weyl function and the wave number}
\begin{theorem}\label{belokolos-enolskii} Let conditions of the
Theorem \ref{its-matveev} are satisfied. Then Weyl function $w(x,P)$ and wave
number $k(P)$  are given by the formulae
\begin{align}
w(x,P)& =\zeta_g\left(\imath x\boldsymbol{e}_g-\boldsymbol{\Omega}
+\int\limits_{(E_{2g+1},0)}^P \mathrm{d}\boldsymbol{h}\right)
-\zeta_g(\imath x\boldsymbol{e}_g -\boldsymbol{\Omega})
-\int\limits_{(E_{2g+1},0)}^P\mathrm{d}r_g. \label{weilzeta}
\end{align}
The wave number or quasi-momentum is
\begin{align}
k(P)&= \left\langle
\zeta_g\left(\imath x\boldsymbol{e}_g-\boldsymbol{\Omega}
+\int\limits_{\infty}^P
\mathrm{d}\boldsymbol{h}\right) -
\zeta_g(\imath x\boldsymbol{e}_g-\boldsymbol{\Omega}) -
 \int\limits_{(E_{2g+1},0)}^P\mathrm{d} r_g\right\rangle.
\end{align}
\end{theorem}
Remark that due to the formula (\ref{zetaformula}) and mentioned
identification $S(x,\lambda)$ as the Bolza polynomial the
quasi-momentum is given as the second kind Abelian integral
(\ref{quasimomentum}). Denote $\mathfrak{a}$-periods of $k(P)$ as
\begin{equation}\mathcal{K}_i
=\oint\limits_{\mathfrak{a}_i}\mathrm{d}\mathcal{K}, \quad
i=1,\ldots, g.
\end{equation}

\subsection{Normalization of the Bloch function}
Different normalizations of the Bloch function are accepted in various
problems of physics. In our case the normalizing constant $C(P)$ is
computed from the condition (\ref{psinorm}). We shall do that on the basis
of the relation (\ref{divadd}).
As the result we obtain the following expression for the
normalizing constant
\begin{equation}
C(P)= \frac{1}{\sqrt{ \langle S(x;\lambda)\rangle  }}.
\label{constant}
\end{equation}
The proof in general case is based on the addition theorem
\cite{bel97a}. Expression (\ref{constant}) was alredy given in
\cite{bbeim94}, Chapter 8. We emphasize that it is valid both for
ergodic and periodic cases.

Further we shall fix the value (\ref{constant}) for the
normalization constant and consider only normalized Bloch functions
$\psi(x,P)$.

\section{Periodic and ergodic finite-gap potentials}
We shall now consider two classes of finite-gap potentials:
periodic and ergodic.
\begin{definition}
We shall call the vector $\boldsymbol{U}=(U_1,\ldots,U_g)^T$,
\begin{equation} \boldsymbol{U} =(2\omega')^{-1}\boldsymbol{e}_g
\end{equation}
as the winding vector
\end{definition}
\begin{definition}
The finite-gap potential is called periodic if the curve $X$  admits
existence of an integer vector $\boldsymbol{n}\in\mathbb{Z}^g $
and real number $U\in \mathbb{R}$ $U\neq 0$ such that
\begin{equation}  \boldsymbol{U}=\boldsymbol{n}U,\label{periodicity}
\end{equation}
otherwise the finite-gap potential is called ergodic.
\end{definition}
We remark that in generic case the finite-gap potential is
ergodic, it is periodic only under special conditions on the curve
presented above.
\subsection{Condition of periodicity of the finite-gap potential } The
direct consequence of the above definition is that the finite-gap
 potential is periodic if and only if period matrix $2\omega'$ satisfy
to the equation for certain vector $\boldsymbol{n}\in\mathbb{Z}^g$ and
real number $U\neq 0$,
\begin{equation}  \omega'\boldsymbol{n}=\frac{1}{2U}
\boldsymbol{e}_g.
\label{percond}  \end{equation} One can easily see that under
condition  (\ref{percond}) the functions in $x$
$g_{kl}(x)=\wp_{kl} \left( \imath x
\boldsymbol{e}_g-\boldsymbol{\Omega} \right)$ are periodic in $x$
with period $\imath/U$. Indeed,
\begin{align*}\wp_{kl} \left(
\imath \left(x+\frac{\imath}{ U} \right)
\boldsymbol{e}_g-\boldsymbol{\Omega} \right) =\wp_{kl} \left(\imath
x \boldsymbol{e}_g-\boldsymbol{\Omega} -2\omega'\boldsymbol{n}
\right) = \wp_{kl} \left(\imath x
\boldsymbol{e}_g-\boldsymbol{\Omega} \right).
\end{align*}
Conditions of periodicity (\ref{percond}) represent $g$
transcendental conditions on the moduli of the curve. But these
conditions are weaker then conditions of double periodicity in $x$ of the
finite-gap potentials. The last conditions are formulated in terms of
period matrix of the curve $\tau$ as follows
\begin{theorem} [{\bf \cite{bbeim94}}]
\label{belokolos-enolskii2}
The finite gap potential $u(x)$ is elliptic function of $x$
if and only if
\begin{itemize}
\item
there exist such homology basis
$\mathcal{B}=(\mathfrak{a}_1,\ldots,\mathfrak{a}_g;\mathfrak{b}_1,\ldots,\mathfrak{b}_g)$
that the period matrix $\tau$ has the form
\begin{equation}
\tau=\left( \begin{array}{ccccc} \tau_{11}&k/N&0&\ldots&0\\
                                  k/N&\ast&\ast&\ast&\ast\\
                                   0&\ast&\ast&\ast&\ast\\
                                   \vdots&\vdots&\vdots&\vdots&\vdots\\
                                    0&\ast&\ast&\ast&\ast
                                    \end{array}  \right),
\label{wreduction1}\end{equation} with $
                                    k\in\{1,\ldots,N-1\}, \quad
                                    N>1,
                                    N\in\mathbb{N}$.
\item
in the homology basis  $\mathcal{B}$ the winding vector
$\boldsymbol{U}$ is of the form
\begin{equation}
\boldsymbol{U}=(\ast,0,\ldots,0)^T.\label{wreduction2}
\end{equation}
\end{itemize}
\end{theorem}
Conditions (\ref{wreduction1}), (\ref{wreduction2}) represent $2g-2$
equations and therefore the  associated double periodic potentials
can be included as particular cases into the set of periodic
potentials. These more particular potentials permit explicit
analytic description and we shall use this circumstance in what
follows. We remark that the reduction theory of Abelian functions to
elliptic functions goes back to K.Weierstrass and A.Poincar\'e and
is exposed in the classical A.Krazer monograpgh\cite{kr03}; the
modern exposition and applications to integrable system are given,
in particular, in \cite{be02a},\cite{be02b}; the Theorem
\ref{belokolos-enolskii2} was recently considered in \cite{gp99},
where an alternative proof is given.
\subsection{Calculation of averages}
The average
\begin{equation}\langle
S(x,\lambda)\rangle=\lambda^g+\sum_{j=1}^g\lambda^{j-1}s_j
\label{spolynom}
\end{equation}
is sensible to commensurability or non-commensurability of the
frequencies (components of the winding vector $\boldsymbol{U}$). In what
follows we shall specify coefficients of the polynomial
(\ref{spolynom}) as $s^{(p)}_j$ or $s^{(e)}_j$ with superscript
$(p)$ taken for periodic case and $(e)$ -- for ergodic, we shall
also provide $\langle S(x,\lambda)\rangle$ with subscripts, $\langle
S(x,\lambda)\rangle_p$ and  $\langle S(x,\lambda)\rangle_e$.
\begin{theorem} [{\bf Averaging in ergodic case}] Let the components of the winding vector $\boldsymbol{U}$
be all non-commensurable. Then the average $\langle
S(x,\lambda)\rangle_e$   is ergodic and given by
\begin{equation}
\langle S(x,\lambda)\rangle_e=\lambda^g+\sum_{j=1}^g\lambda^{j-1}
s_j^{(e)}, \label{average} \end{equation} where
\[ s_j^{(e)}=  \frac{1}{\mathrm{det} 2\omega' }
\mathrm{det}\left( \begin{array}{ccc}
\omega_{1,1}'&\ldots&\omega_{1,g}'\\
\vdots&\ldots&\vdots\\
\omega_{j-1,1}'&\ldots&\omega_{j-1,g}'\\
\eta_{g,1}'&\ldots&\eta_{g,g}'\\
\omega_{j+1,1}'&\ldots&\omega_{j+1,g}'\\
\vdots&\ldots&\vdots\\
\omega_{g,1}'&\ldots&\omega_{g,g}'
\end{array}  \right),\quad j=1,\ldots,g.  \]
In particular, for $g=1$
\begin{align*}
\langle S(x,\lambda)\rangle_e
=\lambda+\frac{\eta'}{\omega'}\end{align*}and for $g=2$
\begin{align*}
\langle S(x,\lambda)\rangle_e& =\lambda^2
+\lambda\frac{\mathrm{det}\left(\begin{array}{cc}\omega_{11}'&\omega_{12}'\\
\eta_{21}'&\eta_{22}'\end{array}    \right)      }
{\mathrm{det}\left(\begin{array}{cc}\omega_{11}'&\omega_{12}'\\
\omega_{21}'&\omega_{22}'\end{array}    \right) }+
\frac{\mathrm{det}\left(\begin{array}{cc}\eta_{21}'&\eta_{22}'\\
\omega_{21}'&\omega_{22}'\end{array}    \right)      }
{\mathrm{det}\left(\begin{array}{cc}\omega_{11}'&\omega_{12}'\\
\omega_{21}'&\omega_{22}'\end{array}    \right) }.
\end{align*}
\end{theorem}
In the case of commensurability we obtain another result for the
average.
\begin{theorem} [{\bf Averaging in periodic case}]  Let the components of the winding vector $\boldsymbol{U}$
be all commensurable. Then the average  $\langle
S(x,\lambda)\rangle_p$  is periodic and given by
\begin{equation}
\langle S(x,\lambda)\rangle_p=z^g+\sum_{j=1}^gz^{j-1} s_j^{(p)},
\label{avper1} \end{equation} where
\begin{equation} s_j^{(p)}=\frac{1}{\mathrm{det} 2\omega' }
 \mathrm{det}\left( \begin{array}{ccc}
\omega_{1,1}'&\ldots&\omega_{1,g}'\\
\vdots&\ldots&\vdots\\
\omega_{g-1,1}'&\ldots&\omega_{g-1,g}'\\
\eta_{j,1}'&\ldots&\eta_{j,g}'
\end{array}  \right),
\quad j=1,\ldots,g.\label{avper2}
\end{equation}
In particular, for $g=1$
\begin{align*}
\langle
S(x,\lambda)\rangle_p=\lambda+\frac{\eta'}{\omega'}
\end{align*}
and for $g=2$
\begin{align*}
\langle S(x,\lambda)\rangle_p& =\lambda^2
+\lambda\frac{\mathrm{det}\left(\begin{array}{cc}\omega_{11}'&\omega_{12}'\\
\eta_{21}'&\eta_{22}'\end{array}    \right)      }
{\mathrm{det}\left(\begin{array}{cc}\omega_{11}'&\omega_{12}'\\
\omega_{21}'&\omega_{22}'\end{array}    \right) }+
\frac{\mathrm{det}\left(\begin{array}{cc}
\omega_{21}'&\omega_{22}' \\
\eta_{11}'&\eta_{12}'\\
  \end{array}    \right)      }
{\mathrm{det}\left(\begin{array}{cc}\omega_{11}'&\omega_{12}'\\
\omega_{21}'&\omega_{22}'\end{array}    \right) }.
\end{align*}
\end{theorem}
We see that for the case of elliptic curve the  averages in ergodic
and periodic case coincide, but for genera bigger than one $\langle
S(x,\lambda)\rangle_p$ and $\langle S(x,\lambda)\rangle_e$ are given
by different expressions. The comprehensive proof of these formulae
will be given elsewhere.

\section{Wannier function}
\begin{definition}
The Wannier function $W_n(x)$ in the $n$-th spectral band,
$n=1,\ldots,g,$ is defined as the following integral of normalized Bloch
function,
\begin{equation}
W_n(x)=\frac{1}{\sqrt{\mathcal{K}_n}}\oint_{\mathfrak{a}_n}
\psi(x,\lambda)\mathrm{d}\mathcal{K}(\lambda),\quad n=1,\ldots,g,
\label{wandef}
\end{equation}
where $\mathcal{K}_n$ are $\mathfrak{a}_n$-periods of the
differential of quasi-momentum (\ref{diffquasimom}).
\end{definition}

Using translation operator one then construct a countable set of
Wannier functions,
\begin{equation}
W_n^{(l)}(x)=\frac{1}{\sqrt{\mathcal{K}_n}}\oint_{\mathfrak{a}_n}
\psi_l(x,\lambda)\mathrm{d}\mathcal{K}(\lambda),\quad
n=1,\ldots,g,l\in\mathbb{Z} \label{wandefcount}
\end{equation}
with
\begin{align}\begin{split}
\psi_l(x,P)&=\frac{1}{\sqrt{ \langle S(x;\lambda)\rangle  }}
\frac{\sigma\left(\int_{\infty}^P\mathrm{d}\boldsymbol{h} -\imath
x\boldsymbol{e}_g+\boldsymbol{\Omega}+2l\boldsymbol{\Omega}'\right)}
{\sigma_{\mathcal{I}} \left(\int_{\infty}^P
\mathrm{d}\boldsymbol{h}\right)
\sigma(\imath x\boldsymbol{e}_g -\boldsymbol{\Omega}-2l\boldsymbol{\Omega}' )}\\
\times& \mathrm{exp}\left\{\imath x\int\limits_{(E_{2g+1},0)}^{P}
\mathrm{d}r_g-(2l\boldsymbol{\Omega}'+ \boldsymbol{\Omega})^T
\int\limits_{(E_{2g+1},0)}^P\mathrm{d}\boldsymbol{r} \right\},
\end{split}
\label{psisigmacount}
\end{align}
where $\boldsymbol{\Omega}'$ is a pure imaginary non-singular even
half-period. In the case of genus $g=1$ the above definitions are
those given in \cite{bes04}.

To complete this definition we must compute periods
(\ref{diffquasimom}). To this regard the following lemma apply.
\begin{lemma}
Denote and $s_k$ are coefficients of the polynomial
\[ \langle S(x,\lambda)\rangle=  \lambda^g+s_g \lambda^{g-1}+\ldots+s_1.
\] Then
\begin{equation}
\left(\begin{array}{c} \mathcal{K}_1\\\vdots\\\mathcal{K}_g
 \end{array}\right)=2\omega^T\left(\begin{array}{c} 2\varkappa_{g,1}-s_1.\\
\vdots\\
2\varkappa_{g,g}-s_g\end{array} \right). \label{perquasimomentum}
\end{equation}
In particular in the case $g=1$, $s_1=\eta'/\omega'$,
$\varkappa_{11}=\varkappa=\eta/2\omega$ and
\begin{equation} \mathcal{K}_1=\frac{\imath\pi}{\omega'}.
\end{equation}
\end{lemma}

Note that in the case $g=1$, the Wannier function (\ref{wandef})
coincides with the classical one derived in \cite{bes04},
\[ W_1(x)=\left( \frac{T}{2\pi} \right)^{1/2} \int_{-\pi/T}^{\pi/T}
\psi(x,P)\mathrm{d} k,
\quad  T=-2\imath \omega'.
 \]

The principal expected properties for the set of Wannier functions,
$W_k(x)$, $k=1,\ldots, g$ is orthogonality
\begin{equation}
\int_{-\infty}^{\infty} \overline{W}_k^{(k')}(x) W_l^{(l')}(x)
\mathrm{d} x =\delta_{kl}\delta_{k'l'}, \quad k,l=1,\ldots, g,\quad
k',l'\in\mathbb{Z}
\end{equation}
and completeness in space of eigenfunctions of the Schr\"odinger
operator with finite-gap quasi-periodic potential. We shall postpone
the discussion of these questions to a future publication.

Since the normalization of Bloch function is different for
periodic and ergodic cases we shall distinguish two kind of
Wannier functions - the classical Wannier functions associated
with periodic potentials and the quasi-periodic Wannier functions
associated with quasi-periodic potentials. Both kinds of  Wannier
functions are defined by the same formula involving Bolza
polynomial in the expression of the normalization constant.
Normalizing constants, however,  are computed in  different ways
for periodic and ergodic cases.

\subsection{Power series for the Wannier function at $|x|\simeq 0$}
\begin{theorem}
The Wannier function of the first lower energy band $W_1(x)$ admits
the following expansion

\begin{equation}W_1(x)=\sum_{p=0}^{\infty}\frac{(-1)^p}{(2p)!}
W_1^{(2p)} x^{2p}.
\end{equation}
Here
\begin{equation}W_1^{(2p)}=\sum_{k=1}M_k q_{pk},
\end{equation}
the quantities $M_k$ are given as $\mathfrak{a}$-cycles of second
kind Abelian integrals, \begin{equation}M_k=
\frac{1}{\sqrt{\mathcal{K}_k}}\oint_{\mathfrak{a}_1}\lambda^k
\sqrt{\frac{\lambda^g+s_g\lambda^{g-1}+\ldots+s_1}{ \prod_{i\in
\{1,\ldots,2g+1 \} -\mathbb{I} }(\lambda-E_i)}}\;\mathrm{d}\lambda
\label{wandiffn}, \quad k=0,\ldots .
\end{equation}
The quantities $q_{kl}$ are coefficients of polynomials in $z$,
$Q_p=\sum_{l=0}^pq_{p,l}z^l$, defined by recurrence
\[ Q_p=\sum_{m=0}^{p-1}\left(\begin{array}{c} 2p\\ 2m-p
\end{array}  \right) \phi_{m-p-1}(z)Q_m
\]
with conditions
\[ \psi_0=\wp_{gg}(\boldsymbol{\Omega})+z,\qquad
\phi_p=\wp_{\begin{array}{c}\underbrace{g\ldots g }\\
\small{2p+2}\end{array}}
(\boldsymbol{\Omega}).
\]

The first few coefficients are
\begin{align*}
W_1^{(0)}&=M_0,\\
W_1^{(2)}&=M_1+2\wp_{gg}(\boldsymbol{\Omega})M_0,\\
W_1^{(4)}&=M_2+4\wp_{gg}(\boldsymbol{\Omega})M_1
+(4\wp_{gg}(\boldsymbol{\Omega})2
+2\wp_{gggg}(\boldsymbol{\Omega}))M_0,\\
W_1^{(6)}&=M_3+6\wp_{gg}(\boldsymbol{\Omega})M_2
+(12\wp_{gg}(\boldsymbol{\Omega})2
+14\wp_{gggg}(\boldsymbol{\Omega}))M_1,\\
&+(2\wp_{gggggg}(\boldsymbol{\Omega})
+28\wp_{gggg}(\boldsymbol{\Omega})\wp_{gg}(\boldsymbol{\Omega})
+8\wp_{gg}(\boldsymbol{\Omega})3)M_0.
\end{align*}
\end{theorem}
Here in its turn
\begin{align*}
\wp_{gg}(\boldsymbol{\Omega})&=\sum_{i\in \mathbb{I} } E_i,\\
\wp_{gggg}(\boldsymbol{\Omega}) &=(6\sum_{i\in \mathbb{I} }
E_i+\alpha_{2k}) \sum_{i\in \mathbb{I} } E_i+4\sum_{i,k\in
\mathbb{I} } E_iE_k,
\end{align*}
where we used (\ref{kdv}).

We remark that the proof of this theorem requires an expansion of
the exponential in $x$ which is a generating function for Schur
polynomials \cite{mac79} and  subsequent integration of these
polynomials (details will be given elsewhere). The integrand can be
expressed recursively in terms of second kind Abelian differentials
described below. In the case of genus one curve and elliptic
periodic potentials these integrals are given as standard
hypergeometric functions \cite{bes04}.

Another remark is that in the course of the proof a reciprocal
hyperelliptic curve $X^{\#}$ of the same genus and dual to the
initial hyperelliptic curve $X$, naturally arises.

\begin{definition} We shall call reciprocal curves $X^{\#}_k$,
$k=1,\ldots,g$ associated to the curve $X$ the set of hyperelliptic curves
of genus $g$  given by the formulae
\begin{align}\label{reciprocal}\begin{split}
{\mu^{\#}}^2
&=  4 (\lambda^g+s_g\lambda^{g-1}+\ldots+s_1)
\prod_{i\in \{1,\ldots,2g+1 \} -\mathbb{I}_k }(\lambda-E_i)\\
&=4(\lambda^{2g+1}+\alpha^{\#}_{2g}\lambda^{2g}+\ldots+\alpha_0^{\#}),
\end{split}\end{align}
\end{definition}
where $\mathbb{I}_k$ is the set of $g$ indices,
$\mathbb{I}_k\subset\{1,\ldots, 2g+1\}$.

We remark that the amplitudes of the Wannier pulses $W_n(0)$,
$n=1,\ldots,g$ are given as periods of meromorphic hyperelliptic
integral of corresponding  reciprocal curves.

To describe canonical differentials on $X_1^{\#}$ it will be
convenient to introduce the set of second kind differentials,
which we shall call the Wannier differentials,
\begin{equation}\mathrm{d}\mathcal{W}_1^{(n)}= \lambda^n\sqrt{\frac{
\lambda^g+s_g\lambda^{g-1}+\ldots+s_1}{ \prod_{i\in
\{1,\ldots,2g+1 \} -\mathbb{I} }(\lambda-E_i)}}\mathrm{d}\lambda
\label{wandiffn}, \quad n=0,\ldots .
\end{equation}
Canonical holomorphic differentials
$\mathrm{d}\boldsymbol{h}^{\#}
=(\mathrm{d}h_1^{\#},\ldots,\mathrm{d}h_g^{\#} )^T$
of the curve  are given as
\begin{equation}  \mathrm{d}h_k^{\#}=\frac{\partial}{\partial s_k}
\mathrm{d}\mathcal{W}^{(0)}
=\frac{\lambda^{k-1}}{\mu^{\#}}\mathrm{d}\lambda,
 \quad k=1,\ldots, g.    \label{holomorphicrec}
\end{equation}
The fact that holomorphic differentials of the reciprocal curve are
gives as derivatives in moduli of meromorphic differentials
resembles principle feature of the algebraic curves appearing in
Seiberg-Witten theory \cite{seiwit94a},\cite{seiwit94b}, see also
\cite{gkmmm95} where various forms of Seiberg-Witten curves were
discussed.

Introduce also canonical second kind differentials
$\mathrm{d}\boldsymbol{r}^{\#}
=(\mathrm{d}r_1^{\#},\ldots,\mathrm{d}r_g^{\#} )^T$  as in
(\ref{meromorphic})
\begin{equation}
\mathrm{d}r_j^{\#}=\sum_{k=j}^{2g+1-j}(k+1-j)\alpha_{k+1+j}^{\#}
\frac{\lambda^k}{4\mu^{\#}} \mathrm{d}\lambda\quad j=1,\ldots,g.
\label{meromorphicrec}
\end{equation}
The first $g$ differentials $\mathrm{d}\mathcal{W}^{(k)}$, $k=0,
\ldots,g-1$ are expressible linearly in terms of $2g$ differentials
$\mathrm{d}\boldsymbol{r}^{\#}$ and $\mathrm{d}\boldsymbol{h}^{\#}$.
Periods of the differentials $\mathrm{d}\mathcal{W}^{(n)}$ with
higher $n$ are also expressible in terms of aforementioned $2g$
differentials due to the de Rham theorem \cite{gh78}.

The last point which we wish to discus here is the asymptotic
expansion of the Wannier function at $x\to +\infty$. To evaluate
this expansion one can use a  variant of the steepest descents
method applicable to the case when the integrand contains a
multiplier vanishing in the saddle point. The following theorem
apply.

\begin{theorem}
\label{wasympt}At $x\to\infty$ the Wannier function of the lower
energy band $[E_1,E_2]$ for the one gap potential has the
following asymptotic expression
\begin{equation} W_1(x)\simeq \mathrm{Re}
\left\{\sqrt{\frac{\mathcal{Q}_0(\lambda_0)}{\mathcal{K}_n}}
     \sqrt{ \prod_{i\in\mathbb{I}} (\lambda_0 -E_i)  } \psi(x,\lambda_0)
 \left[ \frac{\imath}{2\langle S'(x,\lambda_0)\rangle }\right]^{1/4}
\frac{\Gamma\left(\frac34\right)}{x^{\frac34}}\right\},\label{asform}
\end{equation}
where $\lambda_0$ is a solution of the equation
\begin{equation}
\langle S(x,\lambda) \rangle=0, \label{bolzazero}
\end{equation}and $\mathcal{Q}_0(\lambda)$ is given in the Lemma \ref{quotlemma}.
\end{theorem}

The formula for asymptotic expansion naturally generalizes the
result for genus one derived in \cite{bes04} and shows the universal
character of the decreasing low at infinity of the Wannier function:
$\mathrm{exp}\left\{ c_0x\right\} x^{-3/4} $ with phase $c_0x$ in
the saddle point.

\section{Conclusion} In this paper we have defined Wannier functions
associated with permitted zones of quasi-periodic finite gap
potentials. A number of interesting questions remained beyond our
considerations. One question is whether Wannier function introduced
above generate an orthogonal and complete set of in the space of
eigenfunctions of the Schr\"odinger operator. Moreover, a discussion
of  the first non-trivial examples, as well as a comparison between
our analytical expressions for amplitudes and asymptotic behaviors
of Wannier functions with  numerical calculations as done done in
\cite{bes04} for the one gap case, would also be desirable. We plan
to present this material and other related problems in a forthcoming
detailed publication.

\section*{Acknowledgments}\label{sec:Ack}
EDB and VZE wish to thank the Department of Physics of University of
Salerno for a one month financial support, during which part of this
problem was done,  and  to Professors M. Boiti, F. Pempinelli and B.
Prinari for giving them the  opportunity  to participate to the
Conference ``Nonlinear Physics, Theory and Experiment III",
Gallipoli, Italy,  where preliminary results of this paper were
announced. M. S. acknowledges partial financial support from the
MIUR, through the inter-university project PRIN-2003, and from the
Istituto Nazionale di Fisica Nucleare, sezione di Salerno.

\end{document}